\documentclass[a4paper]{jpconf}
\usepackage{graphicx}

\usepackage{epstopdf}
\renewcommand{\e}{\hat\mathbf{e}}

\begin{document}
\title{Numerical simulations of conversion to Alfv\'en waves in solar active regions}

\author{E. Khomenko$^{1}$ and P.~S.~Cally$^2$}

\address{$^1$ Instituto de Astrof\'{\i}sica de Canarias, 38205,
C/ V\'{\i}a L{\'a}ctea, s/n, Tenerife, Spain}
\address{$^2$ School of Mathematical Sciences, Monash University, Clayton, Victoria 3800, Australia}
\ead{khomenko@iac.es}

\begin{abstract}
We study the coupling of magneto-acoustic waves to Alv\'en waves
using 2.5D numerical simulations. In our experiment, a fast
magnetoacoustic wave of a given frequency and wavenumber is
generated below the surface.  The magnetic field in the domain is
assumed homogeneous and inclined. The efficiency of the conversion
to Alfv\'en waves near the layer of equal acoustic and Alfven
speeds is measured calculating their energy flux. The particular
amplitude and phase relations between the oscillations of magnetic
field and velocity help us to demonstrate that the waves produced
after the transformation and reaching upper atmosphere are indeed
Alfv\'en waves. We find that the conversion from fast
magneto-acoustic waves to Alfv\'en waves is particularly important
for the inclination $\theta$ and azimuth $\phi$ angles of the
magnetic field between 55 and 65 degrees, with the maximum shifted
to larger inclinations for lower frequency waves. The maximum
Alfv\'en flux transmitted to the upper atmosphere is about 2--3
times lower than the corresponding acoustic flux.
\end{abstract}

Conversion from fast-mode high-$\beta$ magneto-acoustic waves
(analog of $p$ modes) to slow-mode waves in solar active regions
is relatively well studied both from analytical theories and
numerical simulations (e.g., \cite{Zhugzhda+Dzhalilov1982,
Cally+Bogdan1997, Schunker+Cally2006, Cally2006,
Khomenko+etal2009, Felipe+etal2010a}), see \cite{Khomenko2009} for
a review. In a two-dimensional situation, the transformation from
fast to slow magnetoacoustic modes is demonstrated to be
particularly strong for a narrow range of the magnetic field
inclinations around 20--30 degrees to the vertical.
However, no generalized picture exists so far for conversion from
magneto-acoustic to Alfv\'en waves in a three-dimensional
situation. Studies of this conversion were initiated by Cally \&
Goossens \cite{Cally+Goossens2008}, who found that the
conversion is most efficient for preferred magnetic field
inclinations between 30 and 40 degrees, and azimuth angles between
60 and 80 degrees, and that Alfv\'enic fluxes transmitted to the
upper atmosphere can exceed acoustic fluxes in some cases.
Newington \& Cally \cite{Newington+Cally2010} studied the
conversion properties of low-frequency gravity waves, showing
that large magnetic field inclinations can help transmitting an
important amount of the Alfv\'enic energy flux to the upper
atmosphere.

\begin{figure}[t]
\begin{center}
\includegraphics[width=15cm]{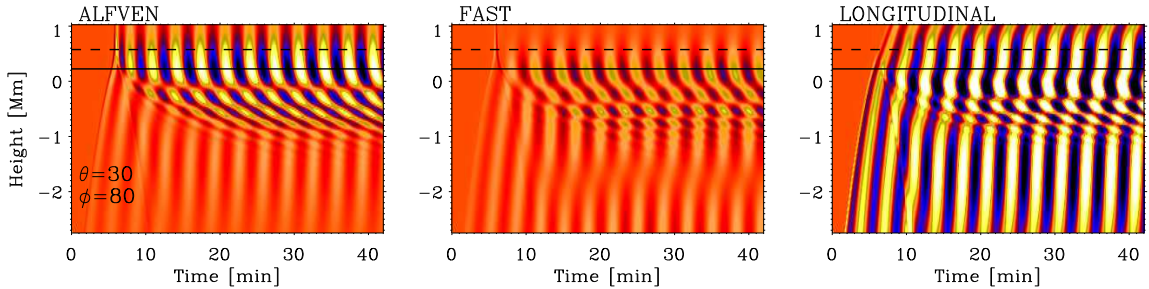}
\end{center}
\caption{\label{fig:modes} Time-height variations of the three
projected velocity components corresponding to $\e_{\rm
perp}$ (Alfven wave, left), $\e_{\rm tran}$ (fast wave,
middle) and $\e_{\rm long}$ (slow wave, right) for $\nu=5$
mHz in a simulation with $B$ inclined by $\theta=30^{\circ}$ and
$\phi=80^{\circ}$. The solid line marks the position $c_S=v_A$, and the dashed
line marks the cut-off layer $\nu=\nu_c\cos\theta$. The colour
scaling is the same in all panels. The amplitudes are scaled with
$\sqrt{\rho_0v_A}$ (first two panels) $\sqrt{\rho_0c_S}$ (last panel).} \vspace{-0.5cm}
\end{figure}

Motivated by these recent studies, here we attack the problem by
means of 2.5D numerical simulations. The purpose of our study is
to calculate the efficiency of the conversion from fast-mode
high-$\beta$ magneto-acoustic waves to Alfv\'en and slow waves in
the upper atmosphere for various frequencies and wavenumbers as a
function of the field orientation. We limit our study to a plane parallel
atmosphere permeated by a constant inclined magnetic field, to
perform a meaningful comparison with the work of Cally \& Goossens
\cite{Cally+Goossens2008}. Numerical simulation will allow generalization 
to more realistic models in our future work.

We numerically solve the non-linear equations of ideal MHD assuming
all vectors in three spatial directions and all derivatives in two
directions (i.e. 2.5D approximation, see
\cite{Khomenko+Collados2006, Khomenko+etal2008}), though
perturbations are kept small to approximate the linear regime. An
acoustic wave of a given frequency and wave number is generated at
$-5$ Mm below the solar surface in a standard model atmosphere
permeated by a uniform inclined magnetic field. The top boundary
of the simulation box is 1 Mm above the surface, and 0.8 Mm above
the layer where the acoustic speed, $c_S$, and the Alfv\'en speed,
$v_A$, are equal. We consider frequencies $\nu=3$ and 5 mHz and
wave numbers $k_X=1.37$ Mm$^{-1}$ and $k_Y=0$. The simulation grid
covers field inclinations $\theta$ from 0$^{\circ}$ to
80$^{\circ}$ and field azimuths $\phi$ from 0$^{\circ}$ to
160$^{\circ}$. The field strength is kept at $B=500$ G. To
separate the Alfv\'en mode from the fast and slow magneto-acoustic
modes in the magnetically dominated atmosphere we use velocity
projections onto three characteristic directions:
\begin{eqnarray}
\label{eq:directions} \e_{\rm long} & = & [\cos\phi
\sin\theta, \, \sin\phi \sin\theta, \, \cos\theta];   \nonumber\\
\e_{\rm perp} & = & [  -  \cos\phi \sin^2\theta \sin\phi, \,
1-\sin^2\theta \sin^2\phi,  \, -  \cos\theta \sin\theta \sin\phi];
\\ \nonumber
\e_{\rm trans} & = & [-\cos\theta, \, 0, \, \cos\phi
\sin\theta].
\end{eqnarray}
To measure the efficiency of conversion to Alfv\'en waves near
and above the $c_S=v_A$ equipartition layer, we calculate acoustic and magnetic energy
fluxes, averaged over time:
\begin{equation}
\label{eq:fluxes} {\bf F_{ac}}  =  \langle p_1{\bf v}_1 \rangle ; \qquad%
{\bf F_{mag}}  =  \langle {\bf B_1}\times({\bf v}_1\times {\bf
B_0})/\mu_0 \rangle .
\end{equation}

Figure \ref{fig:modes} shows an example of the projected
velocities in our calculations as a function of space and time. In
this representation the larger inclination of the ridges mean
lower propagation speeds and vice versa. Note, that by projecting
the velocities, we are able to separate the modes only in the
magnetically dominated atmosphere, i.e. above the solid line in
Fig.~\ref{fig:modes}. The figure shows how the incident fast
mode wave propagates to the equipartition layer and then splits
into several components. The Alfv\'en wave is produced by
mode conversion above 0.2 Mm (left panel) and propagates
upwards with the (rapid) Alfv\'en speed, confirmed by almost vertical
inclination of the ridges. Conversely, the essentially magnetic fast-mode 
low-$\beta$ wave produced in the upper atmosphere (middle panel) is reflected,
and its velocity variations in the upper layers vanish with
height. The (acoustic) slow-mode low-$\beta$ wave escapes to the upper
atmosphere tunnelling over the cut-off layer due to the field
inclination of $\theta=30^{\circ}$. The amplitudes of the velocity
variations of the Alfv\'en wave are comparable to those of the slow
wave.

\begin{figure}[t]
\begin{center}
\includegraphics[width=.35\textwidth]{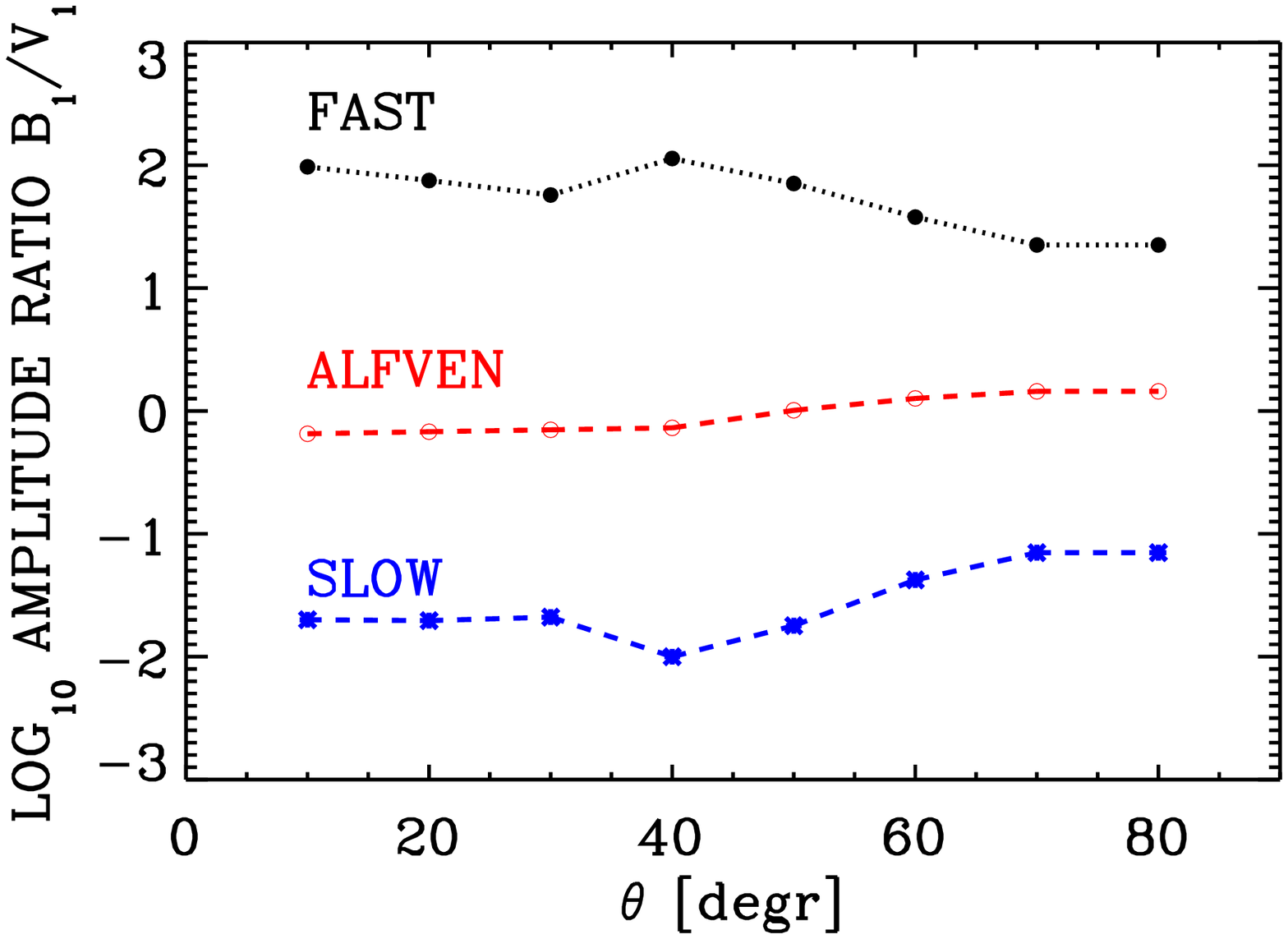}\hspace{3ex}
\includegraphics[width=.35\textwidth]{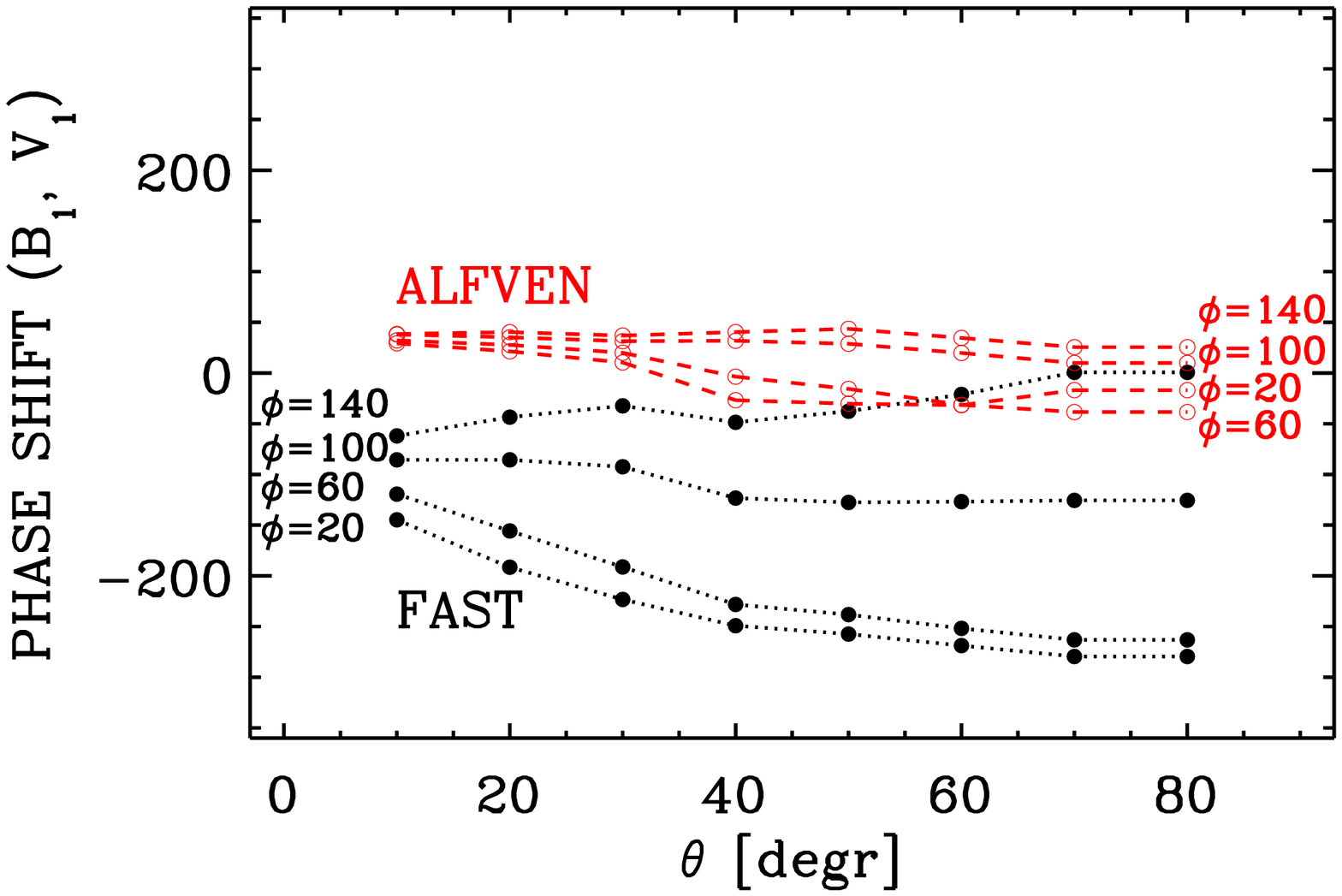}
\end{center}
\caption{\label{fig:phases} Left panel: Log$_{10}$ of the ratio
$B_1$ to $V_1/\sqrt{\mu_0\rho_0}$ for projected velocities and
magnetic field variations, averaged over all $\phi$, as a function
of $\theta$. Black line: fast mode ($\e_{\rm tran}$
projection); red line: Alfv\'en mode ($\e_{\rm perp}$); blue
line: slow mode ($\e_{\rm long}$).  Right panel: phase shift
between the projected variations of $V_1$ and $B_1$, as a function
of $\theta$ for selected $\phi$. Red lines: Alfv\'en mode; black
lines: fast mode.} \vspace{-0.5cm}
\end{figure}


To confirm the Alfv\'en nature of the transformed waves, as
revealed by the projection calculations, we checked the amplitude
and phase relations for all three modes reaching the upper
atmosphere. For the Alfv\'en mode the magnetic field $B_1$ and
velocity variations $V_1$ should be in equipartition (i.e.
$B_1=V_1/\sqrt{\mu_0\rho_0}$), and both magnitudes should
oscillate in phase (see Priest \cite{Priest1984}). Figure
\ref{fig:phases} presents the calculations of the amplitude ratio
$B_1\sqrt{\mu_0\rho_0}/V_1$ and temporal phase shift between $B_1$
and $V_1$, where both velocity and magnetic field variations are
projected in the corresponding characteristic direction for each
mode (Eq.~\ref{eq:directions}). This calculation confirms that,
indeed, for all magnetic field orientations $\theta$ and $\phi$,
the amplitude ratio for the Alfv\'en mode ($\e_{\rm perp}$
projection) is around one (left panel). This is clearly not the
case for the slow and fast modes. For the fast mode, the amplitude
ratio is two orders of magnitude larger, and for the slow mode, it
is two orders of magnitude lower than one. For the Alfv\'en mode
the phase shifts group around zero for all $\phi$, unlike the case
of the fast mode (right panel). We did not calculate the phase
shifts for the slow mode as the variations of the magnetic field
are negligible. Thus, we conclude that the properties of the
simulated Alfv\'en mode separated by the projection correspond to
those expected for a classical Alfv\'en mode.

\begin{figure}[b]
\begin{center}
\includegraphics[width=.3\textwidth]{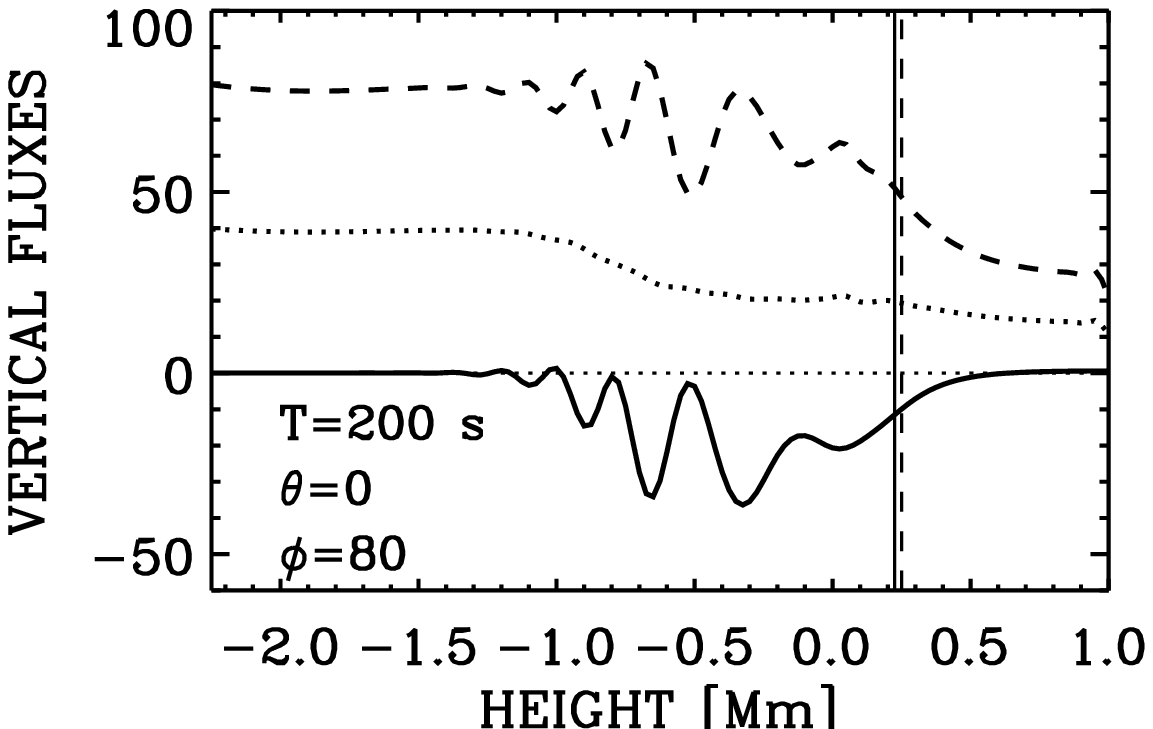}
\includegraphics[width=.3\textwidth]{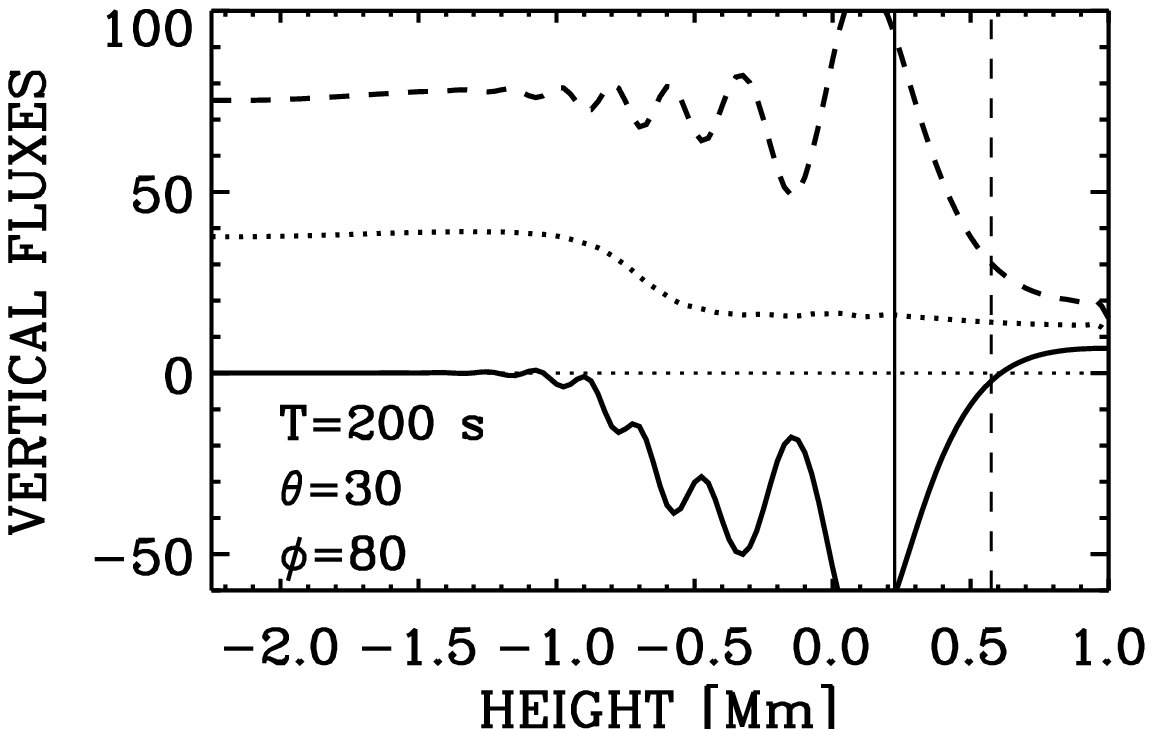}
\includegraphics[width=.3\textwidth]{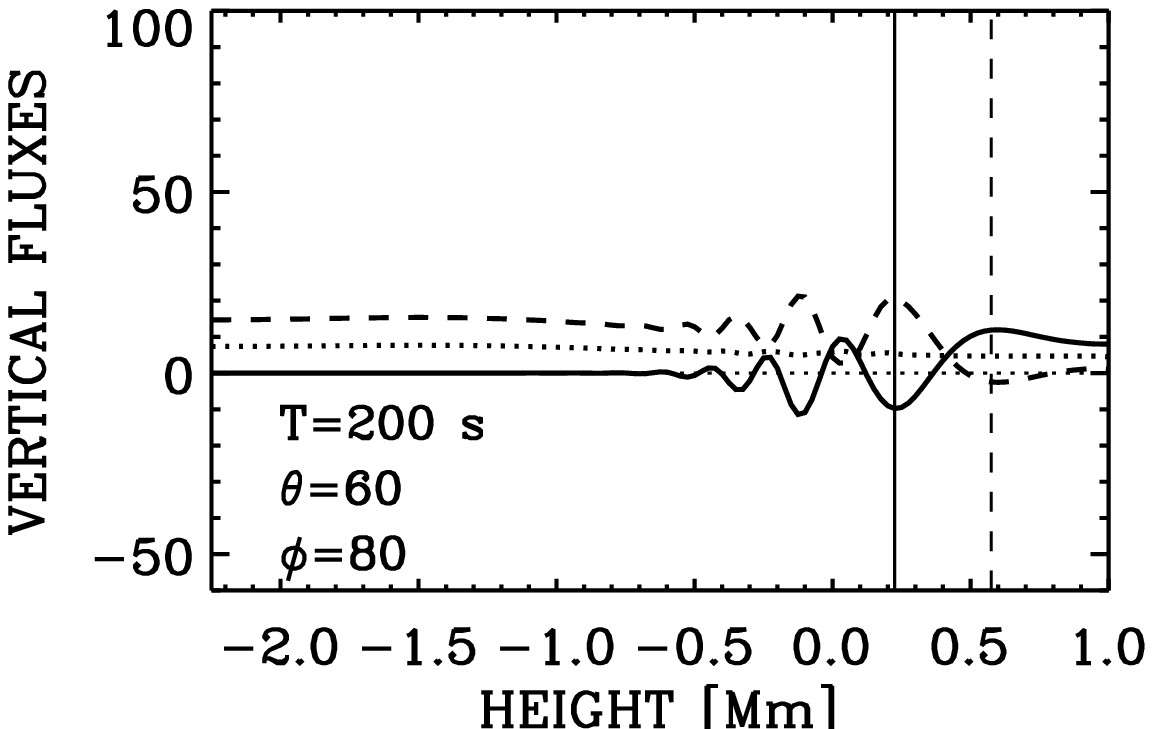}
\end{center}
\caption{\label{fig:fluxes2} Examples of the height dependence of
the magnetic (solid line) and acoustic (dashed line) vertical
fluxes, defined by Eq.~\ref{eq:fluxes}, for $\nu=5$ mHz and
several $\theta$ and $\phi$. Solid vertical line marks the
position $c_S=v_A$, dashed vertical line marks the cut-off layer
$\nu=\nu_c\cos\theta$. } \vspace{-0.5cm}
\end{figure}

An example of the height variations of the acoustic and magnetic
fluxes is given in Figure \ref{fig:fluxes2}. The total vertical
flux (dotted line) is conserved in the simulations except for the
limitations caused by the finite grid resolution not 
resolving slow small-wavelength waves in the deep layers (see
Fig.~\ref{fig:modes}). Both acoustic and magnetic fluxes show
strongest variations near the conversion layer and become constant
above it between 0.5 and 1 Mm height. The fluxes reaching the
upper atmosphere depend crucially on the orientation of the field.
In this example, the acoustic flux decreases with $\theta$ whilst
the magnetic flux increases with $\theta$ and becomes larger than
the acoustic fluxes for $\theta=60^{\circ}$. As the fast wave is
already reflected in the upper atmosphere (see
Fig.~\ref{fig:modes}), the magnetic flux at these heights is due
to the propagating Alfv\'en wave.

\begin{figure}[t]
\begin{center}
\includegraphics[width=6cm]{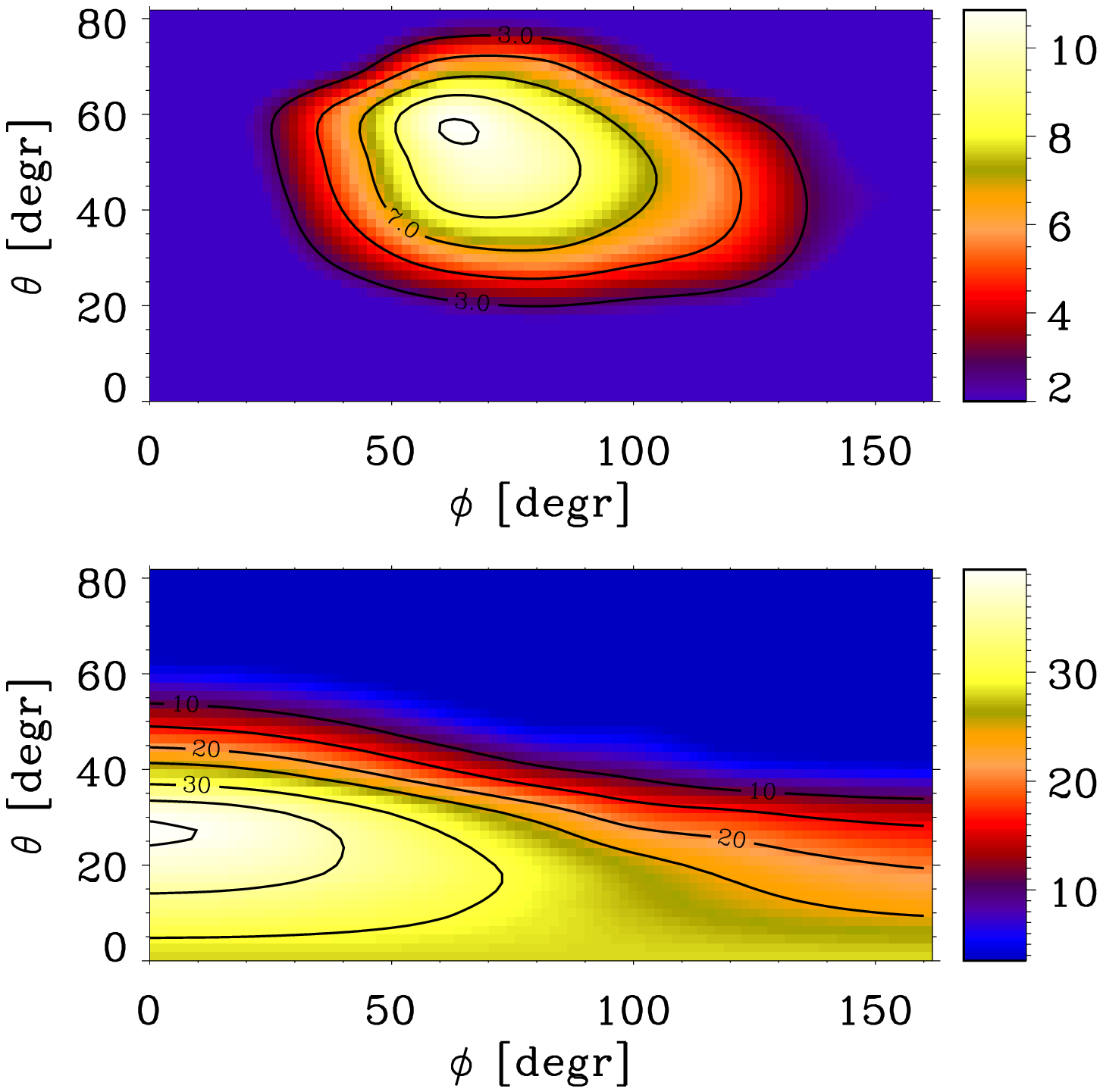}
\includegraphics[width=6cm]{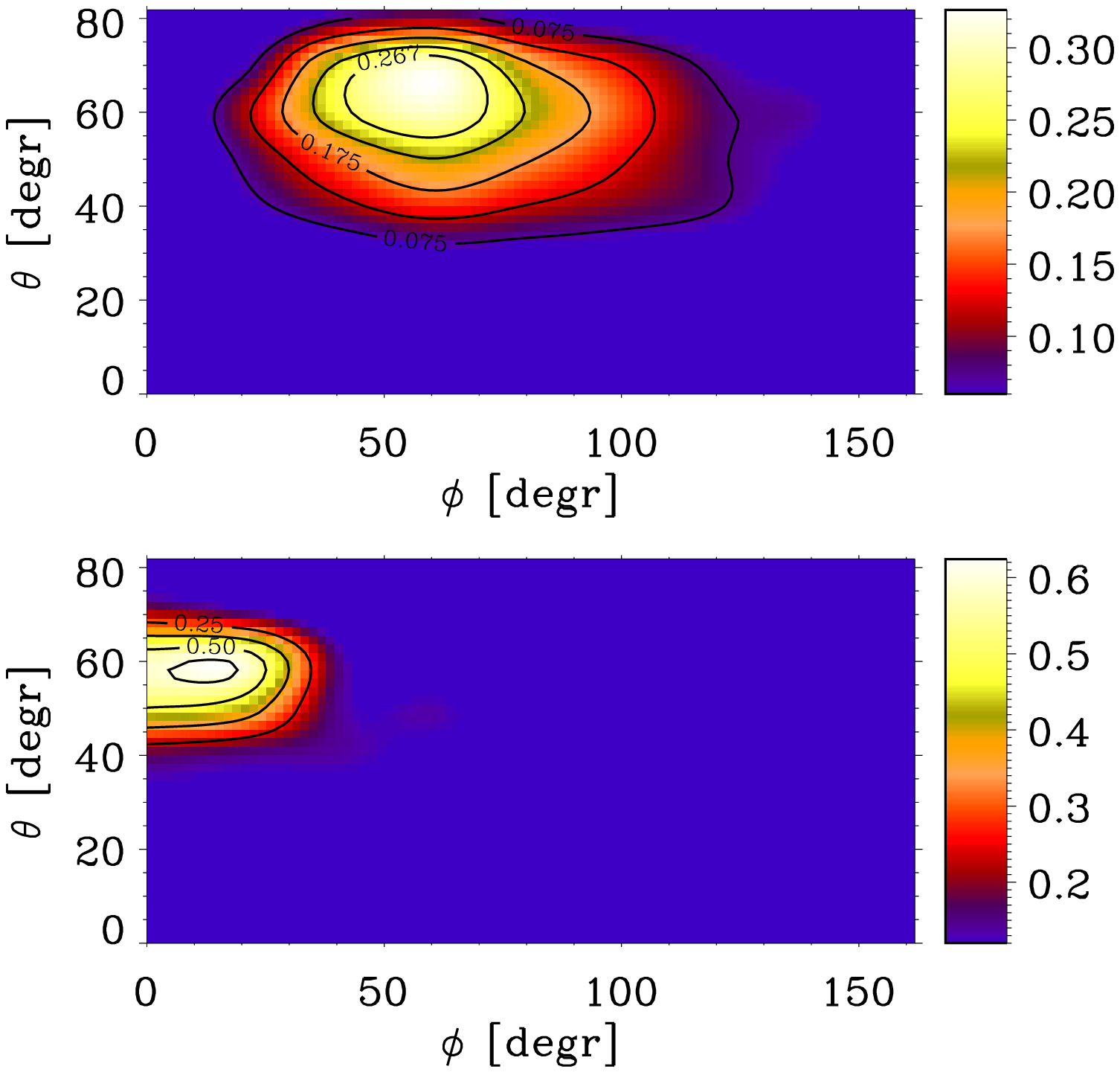}
\end{center}
\caption{\label{fig:fluxes} Vertical fluxes measured at the top of
the atmosphere at 1 Mm for waves with $\nu=5$ mHz (left panels)
and 3 mHz (right panels). Upper panels give magnetic fluxes and
lower panels give acoustic fluxes.  } \vspace{-0.5cm}
\end{figure}

Finally, Figure~\ref{fig:fluxes} gives the time averages of the
vertical magnetic and acoustic fluxes at the top of the atmosphere
as a function of the field orientation. As proven above, the
magnetic flux at 1 Mm corresponds to the Alfv\'en mode. At $\nu=5$
mHz, the maximum of the magnetic flux corresponds to
$\theta=50^{\circ}$ and $\phi=65^{\circ}$. This maximum is shifted
to larger inclinations $\theta=65^{\circ}$ for waves with $\nu=3$
mHz. The presence of the sharp maximum of the Alfv\'enic flux
transmission agrees well with the conclusions made previously by
Cally \& Goossens \cite{Cally+Goossens2008}, though the exact
position of the maximum is shifted to somewhat larger
inclinations. The maximum of the transmitted acoustic flux
corresponds to inclinations $\theta \approx 30^{\circ}$  for
$\nu=5$ mHz waves, and to $\theta \approx 55^{\circ}$ for  $\nu=3$
mHz waves, again, in agreement with previous calculations
\cite{Schunker+Cally2006, Cally+Goossens2008}. The absolute value
of the fluxes is about 30 times lower for 3 mHz compared to 5 mHz.
At some angles the Afv\'en magnetic flux transmitted to the upper
atmosphere is larger than the acoustic flux. However, at angles
corresponding to the maximum of the transmission, the Alfv\'en
flux is 2-3 times lower than the corresponding acoustic flux.

It is important to realize that quantitatively simulating mode
transformation numerically is a challenge, as any
numerical inaccuracies are amplified in such second-order quantities 
as wave energy fluxes. The tests presented in
this paper prove the robustness of our numerical procedure and
offer an effective way to separate the Alfv\'en from magneto-acoustic 
modes in numerical simulations. This will allow us in future to study 
the coupling between magneto-acoustic and Alfv\'en waves in more realistic 
situations resembling complex solar magnetic structures.

\section*{References}

\providecommand{\newblock}{}


\begin{thebibliography}{12}
\expandafter\ifx\csname url\endcsname\relax
  \def\url#1{{\tt #1}}\fi
\expandafter\ifx\csname
urlprefix\endcsname\relax\def\urlprefix{URL }\fi
\providecommand{\eprint}[2][]{\url{#2}}


\bibitem{Zhugzhda+Dzhalilov1982}
Zhugzhda Y~D and Dzhalilov N~S 1982 {\em Astron. Astrophys. \/}
{\bf 112} 16

\bibitem{Cally+Bogdan1997}
Cally P~S and Bogdan T~J 1997 {\em Apj\/} {\bf 486} L67

\bibitem{Schunker+Cally2006}
Schunker H and Cally P~S 2006  {\em MNRAS\/} {\bf 372} 551

\bibitem{Cally2006}
Cally P~S 2006 {\em Phil. Trans. R. Soc. A\/} {\bf 364} 333

\bibitem{Khomenko+etal2009}
Khomenko E, Kosovichev A, Collados M, Parchevsky K and Olshevsky V
2009  {\em Apj\/} {\bf 694} 411

\bibitem{Felipe+etal2010a}
{Felipe} T, {Khomenko} E and {Collados} M 2010 {\em Apj\/} {\bf
719} 357


\bibitem{Khomenko2009}
{Khomenko} E 2009 {\em ASP Conf. Series\/} vol 416 ed. {M~Dikpati,
T~Arentoft, I~Gonz{\'a}lez Hern{\'a}ndez, C~Lindsey, \& F~Hill} p.
31

\bibitem{Cally+Goossens2008}
{Cally} P~S and {Goossens} M 2008 {\em Solar Phys.\/} {\bf 251}
251

\bibitem{Newington+Cally2010}
{Newington} M~E and {Cally} P~S 2010 {\em MNRAS\/} {\bf 402} 386

\bibitem{Khomenko+Collados2006}
Khomenko E and Collados M 2006 {\em Apj\/} {\bf 653} 739

\bibitem{Khomenko+etal2008}
Khomenko E, Centeno R, Collados M and \mbox{Trujillo Bueno} J 2008
{\em Apj\/} {\bf 676}  L85

\bibitem{Priest1984}
Priest E R 1984 {\em Solar magneto-hydrodynamics} Geophysics and
Astrophysics Monographs, Dordrecht: Reidel

\end{thebibliography}
\end{document}